\documentclass[aps,amsmath,amssymb,prl,twocolumn,superscriptaddress]{revtex4}
\usepackage{bm}
\usepackage{epsfig}
\usepackage[usenames]{color}
\usepackage{graphicx}

\newcommand{\el}{{\rm el}}
\newcommand{\inel}{{\rm in}}
\newcommand{\dos}{n}
\newcommand{\ff}{\nu}
\newcommand{\E}{\varepsilon}
\newcommand{\diec}{\epsilon_r}

\newcommand{\av}[1]{\left\langle #1\right\rangle}

\renewcommand{\paragraph}[1]{\textit{#1.---} } %PRL paragraph style: italic & .--- added
\newcommand{\paper}{paper }

\newcommand{\etal}{{\it et al.}}

\begin{document}

\title{Thermoelectric performance of granular semiconductors}

\author{Andreas~Glatz}
\affiliation{Materials Science Division, Argonne National Laboratory, Argonne, Illinois 60439, USA}

\author{I.~S.~Beloborodov}
\affiliation{Department of Physics and Astronomy, California State University Northridge, Northridge, CA 91330, USA}

\date{\today}

\begin{abstract}
We study thermoelectric properties of granular semiconductors with weak tunneling conductance between the grains, $g_t < 1$. We calculate the thermopower and figure of merit taking into account the shift of the chemical potential and the asymmetry of the density of states in the vicinity of the Fermi surface due to n- or p-type doping in the Efros-Shklovskii regime for temperatures less than the charging energy. We show that for weakly coupled semiconducting grains the figure of merit is optimized for grain sizes of order $5$nm for typical materials
and its values can be larger than one. We also study the case of compensated granular semiconductors and show that in this case the thermopower can be still finite, although two to three orders of magnitude smaller than in the uncompensated regime.
\end{abstract}

\maketitle
%%%%%%%%%%%%%%%%%%%%%%%%%%%%%%%%%%%%%%%%%%%%%%%%%%%%%%%%%%%%%%%%%%

A major research area in nano-science and materials physics is the search for highly efficient thermoelectric materials.
Although there has been extensive research efforts in the last several decades, the progress in this quest has been limited until recently.
It was found that for further improvement in efficiency {\it inhomogeneous/granular} thermoelectric semiconductors are especially suited~\cite{Majumdar}.
These materials are now accessible as next generation devices for conversion of thermal to electrical energy and vice versa and technologically important due to the possibility of direct control of the system parameters.
The dimensionless {\it figure of merit}, $ZT=S^2\sigma T/\kappa$, is the preferred measure for the performance or efficiency of thermoelectric materials, where $S$ is the thermopower and $\sigma$ and $\kappa$ the electric and thermal conductivities, respectively~\cite{Rowe,Mahan}.
Recently, $ZT$ values of $2.4$ in layered nanoscale structures~\cite{venka01} at $300K$, and $3.2$ for a bulk semiconductors with nanoscale inclusions~\cite{harman02} at about $600K$  were reported.
These high values of $ZT$ are in the range for applications and therefore call for the development of a comprehensive quantitative description of thermoelectric properties of granular semiconductors, which can serve as basis for a new generation of thermoelectrics.
%We started this task by studying granular metals~\cite{glatz+prb09} and here we will extend our considerations to the case of granular semiconductors.

\begin{figure}[tbh]
\includegraphics[width=0.5\columnwidth]{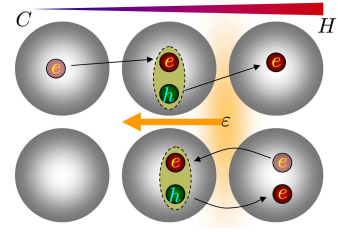}
\caption{(Color online) Sketch of a nanogranular material showing typical electron (e) and hole (h)
transport. In the upper row of grains an inelastic electron tunneling process is shown and in the lower row a co-tunneling loop responsible for the electronic part of the heat transport is presented.
The energy ($\varepsilon$) transport goes from the ''hot'' ($H$) to the ''cold'' ($C$) side.}
\label{fig.model}
\end{figure}

In this \paper we investigate the thermopower $S$ and the figure of merit $ZT$ of
granular semiconductors focusing on the case of weak coupling between the grains, $g_t < 1$, see Fig.~\ref{fig.model}. Each semiconducting nanocrystal is characterized by two energy
scales: (i) the mean energy level spacing $\delta =1/(n a^d)$,
where $n$ is the density of states at the Fermi surface, $a$ is
the grain size, and $d$ is the dimensionality of a grain, and (ii)
the charging energy $E_c = e^2/(\diec a)$ with $\diec$ being the
dielectric constant. In semiconductors the density of states $n$ is of about two
orders of magnitude smaller than that in metals. Thus in
semiconducting dots $\delta$ can be of order of
the charging energy, $\delta \sim E_c$, in contrast to metallic
granular materials where typically $\delta\ll E_c$~\cite{glatz+prb09}. Our considerations are valid for temperatures $T<E_c$.

%{\tt (*) We have: $e^2/(4\pi\epsilon_0 k_B 10nm)=1671$K}

The internal conductance of a grain is taken larger than the inter-grain tunneling
conductance, $g_t$, which controls macroscopic transport properties of the sample~\cite{Beloborodov07}.
In this \paper we consider $g_t<1$, i.e., smaller than the quantum conductance, which is the typical
experimental situation~\cite{venka01,harman02}.

In the case of diagonal (short range) Coulomb interaction, the total probability for an electron tunneling through many grains via
elastic or inelastic co-tunneling can be written as the product, $P  = \prod_{i=1}^{N} P_i$,
of the individual probabilities of single elastic/inelastic co-tunneling
events through single grains with $N=r/a$ is the number of grains.
The probability $P$ is related to the localization length $\xi$ as $P\sim e^{-r/\xi}$.

Semiconducting nanocrystal arrays are described by the Hamiltonian
\begin{equation}
{\cal H}=\sum\limits_i {\cal H}^{(i)}+\sum_{\av{ij},k_i,k_j} \left[t_{ij}{\hat
c}_{k_i}^{(i)\dagger}{\hat c}_{k_j}^{(j)}+\textrm{h.c.}\right] \,, \label{eq.ham}
\end{equation}
where $i,j$ are the grain indexes and the summation in the second term
of the r.h.s. of Eq.~(\ref{eq.ham}) is performed over nearest
neighbors. The term ${\cal H}^{(i)}$ is the Hamiltonian for the
single grain $i$ including the free electron energy and the diagonal Coulomb interaction, and the second term is the tunneling
Hamiltonian between the adjacent grains $i$ and $j$ with $t_{ij}$ being random tunneling matrix elements and ${\hat
c}_{k_i}^{(i)\dagger}$ [${\hat c}_{k_i}^{(i)}$] the creation
[annihilation] operator on the $i$th grain. Due to the large mean energy level spacing in
semiconducting grains $\delta \sim E_c$, only a few terms of the $k$-sums
are important.

In Ref.~\cite{BeloPRB} it was shown that the probability for elastic $P_i^{el}$ and inelastic $P_i^{in}$ co-tunneling through an array of weakly coupled semiconducting grains has the form
\begin{equation}
\label{p_i}
P_i^{\el} = \frac{1}{[\tau \, {\rm max}(\delta, E_c)]^2}, \hspace{0.3cm}
P_i^{\inel} = \frac{e^{-2\delta/T}}{[\tau \, {\rm max}(\delta, E_c)]^2},
\end{equation}
where $\tau$ is the electron escape time from a grain. Thus, for the elastic/inelastic
localization length we obtain
\begin{equation}\label{eq.xiinel}
 \xi^{\el} \sim \frac{a/2}{\ln[\, \tau \, {\rm max} (\delta, E_c)\,]}, \hspace{0.3cm} \xi^{\inel} \sim \frac{a/2}{\ln[\, \tau \, {\rm max} (\delta, E_c)\,] + \delta/T}\,.
\end{equation}
The localization length $\xi^{\el/\inel}$ is related to the characteristic temperature scale
$T_0 = e^2/(\diec \xi^{\el/\inel})$, which is of order $E_c\ln P_i^{-1}$.
\begin{figure}[b]
\includegraphics[width=0.95\columnwidth]{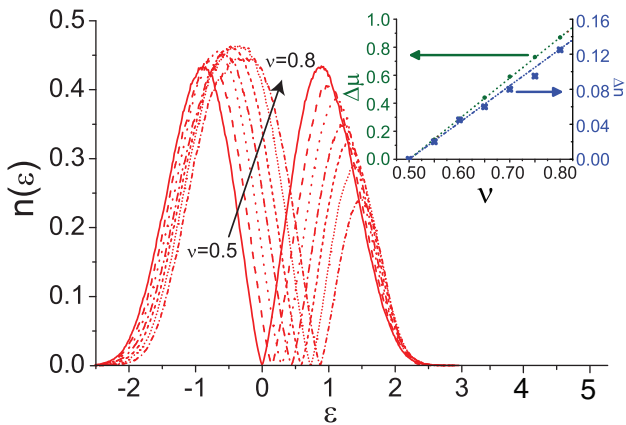}
\caption{(Color online) Simulation results for the density of state $n(\E)$ vs. energy for different filling factors $\ff$, indicated by the arrow across the curves from $\ff=0.5$ to $\ff=0.8$ in steps of $\Delta\ff=0.05$ ($\ff  > 1/2$ corresponds to n-type doped semiconductors). The simulations were done for a 2D Coulomb glass model, simulating the whole weakly coupled granular sample~\cite{glatz+prl07,glatz+jstat08} for a system size of $500^2$. The inset shows the dependence of the shift of the chemical potential $\Delta\mu$ and the asymmetry of the density of states $\Delta\dos$ on $\ff$ (see text). Using these data we extract the numerical coefficients $a_1\approx 2.9$ and $a_2\approx 0.4$ in Eq.~(\ref{S}) by linear fitting.}
\label{fig.dos}
\end{figure}
Below we derive the thermopower, $S$, thermal conductivity, $\kappa$, and figure of merit,
$ZT$ for granular semiconductors using Eq.~(\ref{eq.xiinel}).

One remark is in order: The thermopower $S$ of lightly doped compensated semiconductors was investigated in the past~\cite{Fritzsche71,Zvyagin73,Kosarev74,Parfenov07}. However, all previous studies were concentrated on the Mott variable range hopping (VRH) regime, with conductivity being $\sigma(T) \sim \exp{(T_M/T)^{1/4}}$, where $T_M$ is the Mott temperature~\cite{Mottbook}.
In granular materials the Mott VRH regime is hard to observe.
Indeed, in semiconductors the Efros-Shklovskii (ES)
law~\cite{shklovskii-book,ES} may turn into the Mott behavior with an increase of temperature.
This happens when the typical electron energy $\varepsilon$ involved in a hopping
process becomes larger than the width of the Coulomb gap $\Delta_c$, i.e. when it falls into
the flat region of the density of states where Mott behavior is
expected. To estimate the width of the Coulomb gap $\Delta_c$,
one compares the ES expression for the density of states
$\dos(\Delta_c) \sim (\diec/e^2)^d |\Delta_c|^{d-1}$ with the DOS in
the absence of the long-range part of the Coulomb interaction, $n_0$ ($d = 2,3$ is
the dimensionality of a sample).
Using the condition $\dos(\Delta_c) \sim \dos_0$ we obtain
$\Delta_c = \left( \dos_0 e^{2d}/\diec^d \right) ^{1\over{d-1}  }$. Inserting the value for the bare DOS, $ \dos_0 = 1/E_c \, \xi^d$,
into the last expression we finally obtain $\Delta_c \sim E_c$.
This means that there is no flat region in the DOS for $T<E_c$.

Here we discuss two effects: i) we calculate the thermopower $S$ of granular semiconductors  taking into account the shift of the chemical potential $\Delta \mu = a_1 (\ff - 1/2) T_0$ with $\ff$ being the electron filling factor ($\ff$ is related to the compensation level of semiconductors, here we concentrate on n-type doped semiconductors with $1/2< |\ff| <1$), $a_1$ a dimensionless numerical coefficient, and the asymmetry of the density of states (DOS) $\Delta\dos = a_2 (\ff - 1/2)T_0^{-1}$ with $a_2$ being a numerical constant; ii) we show that even in the absence of the chemical potential shift, $\Delta \mu =0$, and asymmetry of DOS, $\Delta n =0$, the thermopower $S$ is still finite, although small, due to co-tunneling processes.
We start with the former case.

To calculate the thermopower of granular materials in the regime of weak coupling between the grains it is necessary to take into account electrons and holes because the contributions of electrons and holes cancel in the leading order. In general the thermopower is proportional to the average energy transferred by charge carriers and can be written as~\cite{Fritzsche71,Zvyagin73,Kosarev74,Parfenov07}
\begin{equation}
\label{1}
S = - \frac{1}{2e T}\left[ \langle\E - \tilde{\mu}\rangle_e + \langle\E - \tilde{\mu}\rangle_h \right].
\end{equation}
Here the subscripts $e$ and $h$ refer to electrons and holes and $\tilde{\mu} = \mu + \Delta \mu$ is the shifted chemical potential.
The expression in the square brackets of the r.h.s. of Eq.~(\ref{1}) describes the average energy transferred by charge carriers (electron or hole) measured with respect to the shifted chemical potential $\tilde{\mu}$. The average energy in Eq.~(\ref{1}) can be calculated as follows
\begin{equation}
\label{energy}
\langle\E - \tilde{\mu}\rangle_{e/h} = \frac{\int \limits_0^{\infty} d\E\, (\E - \tilde{\mu}) \dos(\E - \tilde{\mu}) f_{e/h}(\E - \tilde{\mu}) e^{-\frac{(\E - \tilde{\mu})^2}{2\Delta^2}}}{\int \limits_0^{\infty} d\E\, \dos(\E - \tilde{\mu}) f_{e/h}(\E - \tilde{\mu}) e^{-\frac{(\E - \tilde{\mu})^2}{2\Delta^2}}}.
\end{equation}
Here  $f_{e/h}(\E )$ is the Fermi function for electrons or holes, $\Delta = \sqrt{T_0 T}$ the typical transfer energy in one hop, and $\dos(\E )$ is energy dependent the DOS.
%One can see that for $\Delta \mu=0$ and $\Delta n =0$, i.e., for $\ff=1/2$ with ES-type DOS, the integral in the nominator of (\ref{energy}) for holes is %equal to the negative one for electrons, which gives zero for the thermopower (\ref{1}).
%Equation~(\ref{energy}) differs in several important ways from previous works~\cite{Fritzsche71,Zvyagin73,Kosarev74,Parfenov07}:
%all considerations in the past were concentrated on the Mott VRH regime with filling factor $\ff = 1/2$.
%In our consideration the average energy, $\langle\E - \tilde{\mu}\rangle_e$ is zero for half filling.
As we will see later,  it is crucial to take into account the asymmetry of the DOS and the shift of the chemical potential in order to obtain a finite result in Eq.~(\ref{1}).

The DOS, $\dos(\E )$ in Eq.~(\ref{energy}) for $\tilde{\mu} - \Delta_c < \E < \tilde{\mu}$ has the following form~\cite{Zvyagin73}
\begin{equation}  \label{dos}
\dos(\E) \propto |\E - \tilde{\mu}|^{d-1}[1 - (\E - \tilde{\mu})\Delta n ]\,,
\end{equation}
and is constant, $n_0$, outside the Coulomb gap region, $\E<\tilde{\mu} - \Delta_c$ and $\E>\tilde{\mu} $, where $\Delta_c \sim E_c$ is the width of the Coulomb gap.
The shift of the chemical potential $\Delta \mu = \tilde{\mu} - \mu$ and the asymmetry of the DOS, $\Delta\dos$, are explicitly defined above Eq.~(\ref{1}).

To support our choice for the expression of the DOS $\dos(\E )$ in Eq.~(\ref{dos}) we numerically compute the DOS for a 2D Coulomb glass model to simulate the whole system of semiconducting grains (see Refs.~\cite{glatz+prl07,glatz+jstat08} for details) at arbitrary filling factor $\ff$ using first principles. The result of the simulations is shown in Fig.~\ref{fig.dos}.
These simulations clearly indicate that for a filling factor $\ff \neq 1/2$, the DOS is asymmetric and the chemical potential is shifted.
Using these results we can identify the dimensionless numerical coefficients $a_1\simeq 2.9$ and $a_2\simeq 0.4$ by a simple linear fit.
We note that $a_1\gg a_2$, thus the contribution to the thermopower caused by $\Delta\mu$ in
Eq.~(\ref{S}) is dominant.

Now, we can calculate Eq.~(\ref{energy}) and the analog contribution for holes using Eq.~(\ref{dos}) for the DOS together with the numerically found values for $a_{1,2}$~\footnote{We expect that the values for $a_{1,2}$ in a 3D Coulomb glass are of the same order as in 2D.}. Finally we derive the expression for the thermopower $S$ of granular semiconductors in the limit of weak coupling between the grains
\begin{equation}
\label{S}
S = - \frac{d\left[\, \Delta\mu +  \Delta\dos T_0 T\right]}{e\, T}=\frac{1/2-\ff}{e}d\left[a_1  \, \frac{T_0}{T} + a_2\right].
\end{equation}
We note that the r.h.s. of Eq.~(\ref{S}) vanishes for filling factor $\ff =1/2$, i.e., for compensated semiconductors.

Equation~(\ref{S}) is valid for temperatures $T < T_0= E_c\ln P^{-1}_i$ and weak coupling between the grains
$g_t < 1$. Under this condition, the electric conductivity $\sigma$
is~\cite{Beloborodov07,BeloPRB}
\begin{eqnarray}
\label{conductivities1}
\sigma  &\simeq& 2e^2 a^{2-d} g_t e^{-\sqrt{T_0/T}}.
\end{eqnarray}

The thermal conductivity $\kappa$ consists of two parts: the electron, $\kappa_e$ and phonon, $\kappa_{ph}$. The phonon contribution $\kappa_{ph}$ at temperatures $T \leq \Theta_D$, where $\Theta_D$ is the Debye temperature is given by~\cite{Ziman, Mahan00}
\begin{eqnarray}
\label{conductivities}
\kappa_{ph} \sim l_{ph}^{2-d} \, T \, \left[T/\Theta_D \right]^{d-1},
\end{eqnarray}
where $l_{ph} = \lambda_F \exp(\Theta_D/[a \, T/\lambda_F])$ is the phonon mean free path in granular semiconductors with $\lambda_F$ being the Fermi length. [For $a\simeq 10$nm, $\lambda_F\simeq 1{\AA}$, $\Theta_D \sim 450K$, one obtains $l_{ph} \simeq 1$nm at $T \simeq 200K$]
\begin{figure}[tbh]
\includegraphics[width=0.8\columnwidth]{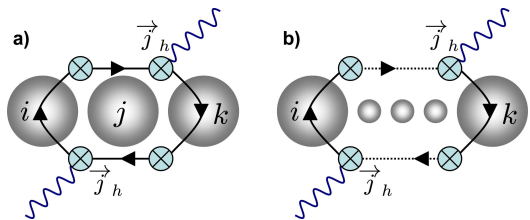}
\caption{(Color online) Diagrams describing lowest (a) and higher (b) order co-tunneling processes (multiple tunneling events are indicated by the dotted lines). The solid lines denote the propagator of electrons. The tunneling vertices are described by the circles. The wavy lines indicate the external coupling to the heat vertices.}
\label{fig.diagram}
\end{figure}
The main contribution to the electric part $\kappa_e$ of the thermal conductivity $\kappa$ appears due to a single closed co-tunneling loop (see Fig.~\ref{fig.model}). An electron executing a co-tunneling loop brings back its charge to the starting grain and hence there is little change in the electrical conductivity and therefore the classical activation term is absent.
However, there is no requirement that the returning electron has exactly the same energy (due to inelastic processes). The leading contribution to $\kappa_e$ is proportional to $g_t^2$ and is depicted in the diagram shown in Fig.~\ref{fig.diagram}a). The analytical result corresponding to this process can be estimated as follows
\begin{equation}
\kappa_e \sim  g_t^2 a^{2-d} \, T \, \frac{e^{-2\delta/T}}{[\tau \, {\rm max}(\delta, E_c)]^2},
\end{equation}
where we used Eq.~(\ref{p_i}) for the inelastic co-tunneling probability $P_i^{\inel}$.
Since at relatively high temperatures the phonon contribution $\kappa_{ph}$ to thermal conductivity is dominant, we can neglect the contribution $\kappa_e$ in the following.

Substituting Eqs.~(\ref{S}) and (\ref{conductivities}) into the expression for the figure of merit $ZT \simeq S^2 \sigma T/\kappa_{ph}$ we obtain the result
\begin{equation}
\label{ZT}
ZT \sim \frac{2d^2 g_t e^{-\sqrt{T_0/T}}\left[ \Delta\mu/T +  \Delta\dos T_0\right]^2}{  (\lambda_F/a)^{2-d} (T/\Theta_D)^{d-1}}e^{(d-2)\frac{\Theta_D\lambda_F}{T a}}\,.
\end{equation}
Here the expressions for $\Delta \mu$ and $\Delta\dos$ are given above Eq.~(\ref{1}).
Using Eq.~(\ref{ZT}) we can calculate the temperature $T^*$ at which $ZT$ has its maximum value, given by the solution of the quadratic equation $T^*=\frac{4}{T_0}\left(\alpha+(d+1)T^*\right)^2$, where $\alpha=(2-d)\Theta_D\lambda_F/a$. In $d=2$ we get $T_{2D}^*=T_0/[4(d+1)]$, while in 3D the existence of a maximum depends on the values of $\alpha$ and $T_0$.

\begin{figure}[tbh]
\includegraphics[width=0.7\columnwidth]{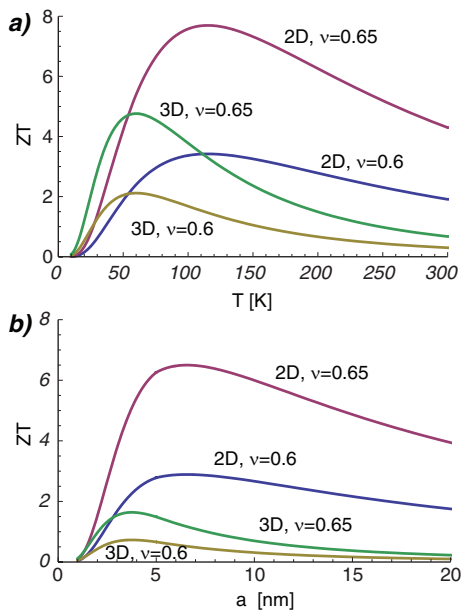}
\caption{(Color online) Plots of the figure of merit $ZT$ vs temperature $T$ (a) and grain size $a$ (b) for two (2D) and three (3D) dimensional samples and two different filling factors $\ff=0.6,0.65$ (n-type doped granular semiconductors). Typical material parameters are given in the text. In a), the grain size is chosen to be $a=5$nm and in b), the temperature is fixed at $200$K. One sees that $ZT$ can be optimized by choosing appropriate grain sizes: For the used parameters, these sizes are about $4$nm in 3D and $6$nm in 2D.}
\label{fig.ZT}
\end{figure}

In Fig.~\ref{fig.ZT} we plotted the figure of merit $ZT$ for a two- and three-dimensional system, using typical parameters for granular semiconductors:
$\lambda_F \simeq 1\AA$, $\Theta_D \simeq 450 {\rm K}$, $T_0 \simeq 2E_c$, $\diec\simeq 4$, and $g_t \simeq 0.1$.
Figure~\ref{fig.ZT}a) shows the temperature dependence of $ZT$ for two different filling factors $\ff=0.6,0.65$ for a grain size of $a\simeq 5$nm (For this size and the above dielectric constant we get $E_c\sim 800$K). We clearly see that the figure of merit can well exceed one. However, we remark that $ZT$ depends inversely proportional on the numerical coefficient of $\kappa_{ph}$, which is assumed to be one here. Figure~\ref{fig.ZT}b) shows the grain size dependence of $ZT$ at fixed temperature $T=200$K. At this temperature for the above system parameters, the figure of merit is optimal for grain sizes $a_{3D}^*\simeq 4$nm in 3D and $a_{2D}^*\simeq 6$nm in 2D.

Now, we concentrate on the compensated regime (filling factor $\ff = 1/2$).
In this case Eq.~(\ref{S}) predicts zero thermopower $S$. Thus, the fundamental question exists
what mechanism may lead to a finite thermopower in this case.
We start our consideration with the fact that the thermopower $S$ can be expressed in terms of the thermoelectric coefficient $\eta$ and the electric conductivity $\sigma$ as $S = \eta/\sigma$.
Thus, to calculate the thermopower $S$ one has to know the thermoelectric coefficient $\eta$. To estimate $\tilde\eta$ we use the diagram shown in Fig.~\ref{fig.diagram}b) (here the tilde indicates the compensated case). Since the dominant contribution to $\tilde\eta$ vanishes due to particle-hole symmetry,
to obtain a nonzero result it is necessary to take into account the fact that the tunneling matrix elements $t$ and the density of states $n$ depend on energy. In the leading order, the corrections to both quantities are proportional to $T/E_F$, where $E_F$ is Fermi energy~\cite{glatz+prb09}. As a result, the thermoelecric coefficient is given by the expression
\begin{equation}
\label{eta}
\tilde\eta \sim e \, a^{2-d}g_t \left[T/E_F \right] e^{-\sqrt{T_0/T}},
\end{equation}
where the temperature scale $T_0$ was defined below Eq.~(\ref{eq.xiinel}). Substituting Eq.~(\ref{conductivities1}) and (\ref{eta}) into the expression for thermopower we obtain
\begin{equation}
\label{S2}
\tilde S \sim (1/e) \, (T/E_F).
\end{equation}
It follows that the thermopower is finite although small, leading to a small figure of merit as well, since $ZT \sim S^2$. Using Eqs.~(\ref{S}) and (\ref{S2}) one can see that the ratio of two thermopowers for compensated
[$\ff =1/2$, Eq.~(\ref{S2})] and for n-type ($\ff >1/2$, Eq.~(\ref{S})) regimes is of order $\tilde S/S\sim T^2/(T_0 E_F) \sim 10^{-3} - 10^{-2} \ll 1$.

In conclusion, we studied thermoelectric properties of granular semiconductors at weak tunneling conductance between the grains, $g_t < 1$. We calculated the thermopower $S$ and figure of merit $ZT$ taking into account the shift of the chemical potential and asymmetry of the density of states for the exemplary case of n-type doping. We showed that the weak coupling between the grains leads to a high thermopower and a low thermal conductivity resulting in relatively high values of the figure of merit. We also discussed the case of compensated (half filling, $\ff = 1/2$) granular semiconductors. We showed that in this regime the thermopower is finite, but rather small.

A.~G. was supported by the U.S. Department of Energy Office of Science under the Contract No. DE-AC02-06CH11357.

\vspace{-0.5cm}

%%%%%%%%%%%%%%%%%%%%%%%%%%%%%%%%%%%%%%%%%%%%%%%%%%%%%%%%%%%%%%%%%%

\end {document}